\documentstyle{lamuphys}
\begin{document}
\title{
{\hspace{9cm}
\rm \small 
HUB-EP-97/79}
\\ \medskip\\
Bosonization in Particle Physics}
\titlerunning{Bosonisation in Particle Physics}
\author{D.Ebert\thanks{Invited talk given at the Workshop ``Field 
Theoretical Tools in Polymer and Particle Physics'', University 
Wuppertal, June 17-19, 1997}}
\institute {Institut f\"ur Physik, Humboldt-Universit\"at zu Berlin,\\
Invalidenstrasse 110, D-10115, Berlin, Germany}
\maketitle
\begin{abstract}
Path integral techniques in collective fields are shown to be a useful 
analytical tool to reformulate a field theory defined in terms of 
microscopic quark (gluon) degrees of freedom as an effective theory 
of collective boson (meson) fields. For illustrations, the path integral 
bosonization approach is applied to derive a (non)linear 
$\sigma$ model from a Nambu-Jona-Lasinio (NJL) quark model. The method 
can be extended to include higher order derivative terms in meson fields 
or heavy-quark symmetries. It is also approximately applicable to QCD. 
\end{abstract}
\section{Introduction}
In this lecture, I want to demonstrate the powerfulness of the path 
integral 
approach in collective fields for the bosonization of quark models 
containing (effective) 4-quark interactions [1]. For 
illustrations, let 
me consider NJL type of models [2-4] with {\em local} quark 
interactions $\sim G\left(\bar q\Gamma q\right)^2$ representing 
a relativistic version of the superconductor BCS theory [5,6]. 
These models lead to a gap equation for a dynamical quark mass 
signalling the dynamical breakdown of chiral symmetry (DBCS). 
Furthermore, the collective field of Cooper pairs of the superconductor 
is now replaced by collective meson fields of 
$\left(q\tilde q\right)$-bound states. 

To be more explicit, let me 
consider the generating functional ${\cal Z}$ of the NJL model defined 
by a path 
integral of the exponential of the corresponding action over quark fields 
$q$ as the underlying microscopic degrees of freedom. Path integral 
bosonization then means to transform this generating functional into an 
integral of the exponential of an effective meson action where the new 
collective integration variables $\sigma, \vec\pi, \vec\rho,...$ denote the 
observable meson fields,  

\begin{equation}  
{\cal Z}=\int D\mu(q)\E^{\I\int {\cal L}_{\rm NJL}\D^4 x}
\stackrel{I}{\Longrightarrow}
\int D\mu\left(\sigma,\vec\pi, \vec\rho,...\right)\E^{\I\int 
{\cal L}_{\rm Eff.}\D^4 x},
\end{equation}
with $D\mu (q)=Dq D\bar q,~ D\mu\left(\sigma, \vec\pi, \vec\rho,...\right)=
D\sigma 
D\vec\pi D\vec\rho...$ being the respective integration measures of fields.

The basic ingredient of the path integral bosonization (1) is the use of 
the Hubbard-Stratonovich transformation [7,8] which replaces the (effective) 
4-quark interactions of NJL models by a Yukawa-type coupling of quarks 
with collective meson fields $\phi_i=\left(\sigma, \vec\pi, 
\vec\rho_\mu,...\right)$. After this the primary path integral 
over quark fields 
on the L.H.S. of (1) becomes Gaussian resulting in a quark determinant 
containing meson fields. Further important steps are: 

\begin{enumerate}
\begin{enumerate}
\item

the use of the loop expansion of the quark determinant in powers of meson 
fields 
\item

the evaluation of the resulting Feynman diagrams in the low-momentum 
region  

\item

the limit of large numbers of colours, $N_{\rm c}\to\infty$, $GN_{\rm c}$ 
fixed, $G$ being the 4-quark coupling constant in order to apply a 
saddle point approximation to the integration over meson fields.

\end{enumerate}
\end{enumerate}
Quark loop diagrams emitting two $\phi$-fields then generate in the 
low-momentum (two-derivative) approximation kinetic and mass terms 
of mesons. Finally, quark loops emitting $n>2$ meson fields lead to 
meson interactions with effective small coupling constants $g_n\sim
O\left(\left(\frac{1}{\sqrt{N_{\rm c}}}\right)^{n-2}\right)$ allowing 
for a modified perturbation theory in terms of meson degrees of freedom.     

Notice that the NJL quark Lagrangian incorporates the global chiral flavour 
symmetry of Quantum Chromodynamics (QCD) as well as its explicit and 
dynamical breaking. The equivalent effective meson theory on the R.H.S. 
of (1) just reproduces this symmetry breaking pattern at the meson level. 
As mentioned above, masses and coupling constants of collective mesons 
are now calculable from quark loop diagrams and expressed by the 
parameters of the underlying quark theory (including a loop momentum 
cut-off $\Lambda$). The final aim is, of course, bosonization of QCD, 
i.e. the transformation 

\begin{equation}
{\cal Z}=\int D\mu\left(q, G_\mu\right)\E^{\I\int {\cal L}_{\rm QCD}\D^4x}
\stackrel{II}{\Longrightarrow}\int D\mu\left(\sigma,\vec\pi,\vec\rho,...
\right)
\E^{\I\int {\cal L}_{\rm Eff.}\D^4 x}, 
\end{equation}
where $q, G_\mu$ denote quark and gluon fields (ghost fields are not 
shown explicitly). In order to begin with an effective 4-quark interaction 
as intermediate step, one first has to integrate away the gluon 
(ghost) fields 
on the L.H.S. of (2). This can only be done exactly for space-time 
dimensions $D=2$ in the light-cone gauge, $G_{-}=\frac12\left(G_0-G_1
\right)=0$, where all self-interactions of gluon fields vanish. The 
resulting expression contains an effective {\em nonlocal}  
$\mbox{current}
\times \mbox{current}$ quark interaction with a known gluon propagator 
which then can be bosonized by introducing {\em bilocal} meson fields 
$\phi\left(x,y\right)\sim \bar q(x)\Gamma q(y)$, employing the limit of 
large 
numbers of colours $N_{\rm c}$ [9]. Clearly, the analogous bosonization 
of four-dimensional QCD is more complicate. It requires some additional 
approximations: first a truncation of multilocal quark current 
interactions retaining only the bilocal two-current term and secondly 
a modelling of the unknown nonperturbative gluon propagator (Cf. Fig. 1a) 
(see [1,10-13]). 
As a final result one derives an effective chiral Lagrangian describing 
the low-energy interactions of observable mesons including nontrivial 
meson form factors.

\begin{figure}[htb]

\def\emline#1#2#3#4#5#6{%
       \put(#1,#2){\special{em:moveto}}%
       \put(#4,#5){\special{em:lineto}}}
\def\newpic#1{}
\unitlength=0.70mm
\special{em:linewidth 0.4pt}
\linethickness{0.4pt}
\begin{center}
\begin{picture}(130.00,60.00)
\emline{40.00}{60.00}{1}{50.00}{50.00}{2}
\emline{50.00}{50.00}{3}{60.00}{60.00}{4}
\put(50.00,48.00){\oval(4.00,4.00)[l]}
\put(50.00,44.00){\oval(4.00,4.00)[r]}
\put(50.00,40.00){\oval(4.00,4.00)[l]}
\put(50.00,30.00){\oval(4.00,4.00)[r]}
\put(50.00,26.00){\oval(4.00,4.00)[l]}
\put(50.00,22.00){\oval(4.00,4.00)[r]}
\emline{50.00}{20.00}{5}{40.00}{10.00}{6}
\emline{50.00}{20.00}{7}{60.00}{10.00}{8}
\emline{90.00}{50.00}{9}{120.00}{20.00}{10}
\emline{120.00}{50.00}{11}{90.00}{20.00}{12}
\put(105.00,35.00){\circle*{3.00}}
\emline{112.00}{42.00}{13}{109.00}{42.00}{14}
\emline{112.00}{42.00}{15}{112.00}{39.00}{16}
\emline{109.00}{31.00}{17}{109.00}{28.00}{18}
\emline{112.00}{31.00}{19}{109.00}{31.00}{20}
\emline{101.00}{28.00}{21}{98.00}{28.00}{22}
\emline{98.00}{31.00}{23}{98.00}{28.00}{24}
\emline{101.00}{39.00}{25}{101.00}{42.00}{26}
\emline{101.00}{39.00}{27}{98.00}{39.00}{28}
\put(74.00,35.00){\makebox(0,0)[cc]{low energy}}
\put(67.00,32.00){\vector(1,0){14.00}}
\put(35.00,55.00){\makebox(0,0)[cc]{$q$}}
\put(65.00,55.00){\makebox(0,0)[cc]{$q$}}
\put(35.00,15.00){\makebox(0,0)[cc]{$\tilde q$}}
\put(65.00,15.00){\makebox(0,0)[cc]{$\tilde q$}}
\put(50.00,35.00){\circle*{6.00}}
\emline{46.00}{54.00}{29}{46.00}{57.00}{30}
\emline{57.00}{57.00}{31}{57.00}{54.00}{32}
\emline{54.00}{57.00}{33}{57.00}{57.00}{34}
\emline{46.00}{54.00}{35}{43.00}{54.00}{36}
\emline{54.00}{16.00}{37}{57.00}{16.00}{38}
\emline{54.00}{16.00}{39}{54.00}{13.00}{40}
\emline{44.00}{14.00}{41}{44.00}{17.00}{42}
\emline{44.00}{14.00}{43}{47.00}{14.00}{44}
\put(105.00,45.00){\makebox(0,0)[cc]{$\frac{G}{2}$}}
\put(85.00,45.00){\makebox(0,0)[cc]{$q$}}
\put(125.00,45.00){\makebox(0,0)[cc]{$q$}}
\put(85.00,25.00){\makebox(0,0)[cc]{$\tilde q$}}
\put(125.00,25.00){\makebox(0,0)[rb]{$\tilde q$}}
\put(130.00,35.00){\makebox(0,0)[cc]{$=\quad {\cal L}_{\rm NJL}^{\rm int.}$ }}
\put(50.00,0.00){\makebox(0,0)[cb]{a)}}
\put(105.00,0.00){\makebox(0,0)[cb]{b)}}
\end{picture}
\end{center}
\caption[ ]{a-b) Low-energy approximation of a nonlocal current~$\times$
current interaction with nonperturbative gluon propagator (a), by a local
Nambu-Jona-Lasinio type interaction (b).}
\end{figure}
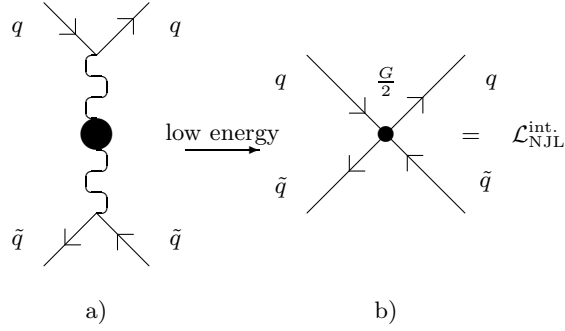

Nevertheless, as anticipated in (1), also simpler 
{\em local} NJL type of models are expected to yield a reasonable low-energy 
description of the chiral sector of ${\rm QCD}_4$. Indeed, discarding 
the complicated question of the unknown structure of nonperturbative 
gluon and quark propagators (related to confinement), one can try 
to approximate the nonperturbative gluon propagator in the region of low 
energies by a universal constant $G$ leading to a {\em local} NJL type 
of ${\rm current}\times {\rm current}$ interaction (Cf. Fig. 1). (For 
interesting nonperturbative applications of the above approach the 
reader is further referred to the nuclear many-body problem [10,14] and 
the Hubbard model [15,16].) 
 
In conclusion, let us notice that for two space-time dimensions 
there exists a different realization of the bosonization idea based 
on an explicit construction of fermionic fields in terms of bosonic 
fields due to Mandelstam [17]. This then allows one to replace, at 
the operator level, quark bilinears by bosonic fields, e.g. 

$$\bar\psi\gamma_\mu\psi\sim -\pi^{-\frac12}\varepsilon_{\mu\nu}
\partial^\nu\phi,~~\bar\psi\psi\sim M\cos\left(2\sqrt{\pi}\phi\right),$$
and to prove the equivalence of a given fermionic model (e.g. Thirring 
model) with a corresponding bosonic model (e.g. Sine-Gordon model). 
Using in particular Witten's non-Abelian bosonization rules [18] one 
has derived in the strong coupling limit (which is contrary to the 
weak coupling limit, $GN_{\rm c}$ fixed, for $N_{\rm c}\to\infty$, of 
the above described path integral approach) a low-energy effective 
action from ${\rm QCD}_2$ [19]. The lecture of T.Giamarchi at this 
Workshop [20] discusses just the application of this kind of operator 
bosonization in Condensed Matter Physics.

In the next section I shall now describe the path integral bosonization 
of the NJL model along line $I$ closely following the original papers 
[3,4].

\section{NJL Model and $\sigma$ Model}
\subsection{Linear $\sigma$ Model}
Let us consider the following NJL Lagrangian with a global symmetry   
\\
$\left[U(2)\times U(2)\right]\times SU(N_{\rm c})$ [3] 

\begin{equation}
{\cal L}_{{\rm NJL}}=\bar q\left(\I\hat\partial-m_0\right) q
+\frac{G}{2}\left[\left(\bar q q\right)^2+\left(\bar q\I\gamma_5
\vec \tau q\right)^2\right], 
\end{equation}
where $q$ denotes a quark spinor with colour and spinor indices, 
$\tau_i$'s are the generators of the flavour group $U(2)$, 
and $G$ 
is a universal quark coupling constant with dimension 
$\mbox{mass}^{-2}$. Note that the $\bar q q$-combinations in (3) are 
colour singlets, and we have admitted an explicit chiral 
symmetry-breaking current quark mass term $-m_0\bar q q$. The 
integration over the quark fields in the generating functional 
${\cal Z}$ of the NJL model (given by the L.H.S. of (1)) can easily 
be done after bi-linearizing the four-quark interaction terms with 
the help of colour-singlet collective meson fields $\sigma,\vec \pi$.
To this end, we use the Hubbard-Stratonovich transformation [7,8]

$$\exp\Biggl\{\I\int\frac{G}{2}\left[\left(\bar q q\right)^2+
\left(\bar q \I\gamma_5\vec \tau q\right)^2\right]\D^4x\Biggr\}=$$

\begin{equation}
={\cal N}\int D\sigma D\pi_i\exp\Biggl\{\I\int\left[-\frac{1}{2G}
\left(\sigma^2+\vec \pi^2\right)-\bar q \left(\sigma+\I\gamma_5
\vec\tau\cdot\vec\pi\right)q\right]\D^4x\Biggr\},
\end{equation}        
which replaces the 4-quark interaction by a Yukawa coupling with 
collective fields $\sigma,\vec \pi$, Cf. Fig.2.

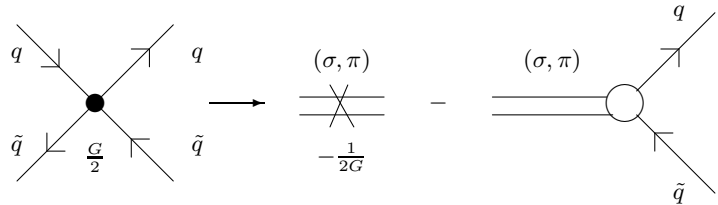
\begin{figure}[htb]

\def\emline#1#2#3#4#5#6{%
       \put(#1,#2){\special{em:moveto}}%
       \put(#4,#5){\special{em:lineto}}}
\def\newpic#1{}
\unitlength=0.8mm
\special{em:linewidth 0.4pt}
\linethickness{0.4pt}
\begin{picture}(136.00,32.00)
\emline{20.00}{30.00}{1}{46.00}{04.00}{2}
\emline{46.00}{30.00}{3}{20.00}{04.00}{4}
\put(33.00,17.00){\circle*{3.00}}
\emline{27.00}{23.00}{5}{27.00}{26.00}{6}
\emline{42.00}{26.00}{7}{39.00}{26.00}{8}
\emline{42.00}{26.00}{9}{42.00}{23.00}{10}
\emline{25.00}{09.00}{11}{25.00}{12.00}{12}
\emline{28.00}{09.00}{13}{25.00}{09.00}{14}
\emline{39.00}{11.00}{15}{39.00}{08.00}{16}
\emline{39.00}{11.00}{17}{42.00}{11.00}{18}
\emline{27.00}{23.00}{19}{24.00}{23.00}{20}
\put(20.00,25.00){\makebox(0,0)[cc]{$q$}}
\put(50.00,25.00){\makebox(0,0)[cc]{$q$}}
\put(20.00,10.00){\makebox(0,0)[cc]{$\tilde q$}}
\put(50.00,10.00){\makebox(0,0)[cc]{$\tilde q$}}
\put(33.00,08.00){\makebox(0,0)[cc]{$\frac G2$}}
\put(52.00,17.00){\vector(1,0){9.00}}
\emline{67.00}{18.00}{21}{81.00}{18.00}{22}
\emline{67.00}{15.00}{23}{81.00}{15.00}{24}
\emline{72.00}{20.00}{25}{76.00}{13.00}{26}
\emline{75.00}{20.00}{27}{72.00}{13.00}{28}
\put(74.00,24.00){\makebox(0,0)[cc]{$\left(\sigma, \vec\pi\right)$}}
\put(74.00,08.00){\makebox(0,0)[cc]{$-\frac 1{2G}$}}
\emline{99.00}{18.00}{29}{118.00}{18.00}{30}
\emline{99.00}{15.00}{31}{119.00}{15.00}{32}
\put(121.00,17.00){\circle{6.32}}
\emline{123.00}{19.00}{33}{136.00}{32.00}{34}
\emline{123.00}{15.00}{35}{136.00}{02.00}{36}
\emline{130.00}{26.00}{37}{127.00}{26.00}{38}
\emline{130.00}{26.00}{39}{130.00}{23.00}{40}
\emline{126.00}{12.00}{41}{129.00}{12.00}{42}
\emline{126.00}{12.00}{43}{126.00}{09.00}{44}
\put(109.00,23.00){\makebox(0,0)[cb]{$\left(\sigma, \vec\pi\right)$}}
\put(131.00,32.00){\makebox(0,0)[rc]{$q$}}
\put(130.00,02.00){\makebox(0,0)[cc]{$\tilde q$}}
\put(90.00,17.00){\makebox(0,0)[cc]{$-$}}
\end{picture}
\caption[ ]{Graphical representation of the integral identity (4).}
\end{figure}
\noindent
Inserting (4) into the L.H.S. of (1) leads to the intermediate 
result

\begin{equation}
{\cal Z}=\int D\sigma D\pi_i\int Dq D\bar q \E^{\I\int 
{\cal L}_{{\rm NJL}}^{{\rm qM}}\D^4 x}
\end{equation}
with the semi-bosonized meson-quark Lagrangian

\begin{equation}
{\cal L}_{{\rm NJL}}^{{\rm qM}}=-\frac{1}{2G}\left(\left(
\sigma-m_0\right)^2+\vec \pi^2\right)+\bar q\left(\I\hat
\partial-\sigma-\I\gamma_5\vec\tau\cdot\vec\pi\right)q,
\end{equation}
where we have absorbed the current quark mass in the $\sigma$ 
field by a shift $\sigma\to\sigma-m_0$ of the integration 
variable. The Gaussian integral over the Grassman variable 
$q$ in (5) can easily be performed leading to the fermion 
determinant 

\begin{equation}
\det S^{-1}=\exp N_{\rm c} {\rm Tr}{\,}\ln S^{-1}=\exp \I N_{\rm c}
\int -\I{\,}{\rm tr}{\,}\left(\ln{\,}S^{-1}\right)_{(x,x)}\D^4x
\end{equation}
where

\begin{equation}
S^{-1}(x,y)=\left[\I\hat\partial_x-\sigma(x)-\I\gamma_5\vec 
\tau\cdot\vec\pi(x)\right]\delta^4(x-y)
\end{equation}
is the inverse quark propagator in the presence of collective 
fields $\sigma,\vec\pi$ and the trace tr in (7) runs over 
Dirac and isospin indices. Note that the factor $N_{\rm c}$ results 
from the colour trace. Combining (5)-(7) we obtain 

\begin{equation}
{\cal Z}=\int D\sigma D\pi_i\E^{\I\int {\cal L}_{\rm Eff.}
\left(\sigma,\vec \pi\right)\D^4x},
\end{equation}

\begin{equation}
{\cal L}_{\rm Eff.}\left(\sigma,\vec \pi\right)=-\frac{1}{2G}
\left(\left(\sigma-m_0\right)^2+\vec\pi^2\right)-\I{\,}N_{\rm c}
{\rm tr}{\,}\ln\left(\I\hat\partial-\sigma-\I\gamma_5\vec\tau
\cdot\vec\pi\right)_{(x,x)}.
\end{equation}
Let us analyze (9) in the limit of large $N_{\rm c}$ with $GN_{\rm c}$ 
fixed where one can apply the saddle point approximation. 
The stationary point $\sigma=\left<\sigma\right>_0\equiv m, 
\vec \pi=0$ satisfies the condition

\begin{equation}
\left.\frac{\delta {\cal L}_{{\rm Eff.}}}{\delta\sigma}
\right|_{\left<\sigma\right>_0,\vec\pi=0}=0,
\end{equation}
which takes the form of the well-known Hartree-Fock gap 
equation determining the dynamical quark mass $m$

\begin{equation}
m=m_0+8mGN_{\rm c}I_1(m),
\end{equation}
\begin{equation}
I_1(m)=\I\int\limits_{}^\Lambda\frac{\D^4k}
{\left(2\pi\right)^4}\frac{1}{k^2-m^2}
\end{equation}
with $\Lambda$ being a momentum cut-off which has to be held 
finite (see Fig.3).

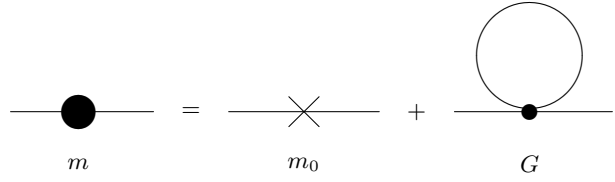
\begin{figure}[htb]

\def\emline#1#2#3#4#5#6{%
       \put(#1,#2){\special{em:moveto}}%
       \put(#4,#5){\special{em:lineto}}}
\def\newpic#1{}

\unitlength=1mm
\special{em:linewidth 0.4pt}
\linethickness{0.4pt}
\begin{picture}(101.00,40.00)
\emline{21.00}{15.00}{1}{40.00}{15.00}{2}
\put(30.00,15.00){\circle*{4.47}}
\put(45.00,15.00){\makebox(0,0)[cc]{=}}
\emline{50.00}{15.00}{3}{70.00}{15.00}{4}
\emline{58.00}{17.00}{5}{62.00}{13.00}{6}
\emline{62.00}{17.00}{7}{58.00}{13.00}{8}
\put(75.00,15.00){\makebox(0,0)[cc]{+}}
\emline{80.00}{15.00}{9}{101.00}{15.00}{10}
\put(90.00,22.50){\circle{15.00}}
\put(90.00,15.00){\circle*{2.00}}
\put(30.00,8.00){\makebox(0,0)[cc]{$m$}}
\put(60.00,8.00){\makebox(0,0)[cc]{$m_0$}}
\put(90.00,8.00){\makebox(0,0)[cc]{$G$}}
\end{picture}

\caption{Graphical representation of the gap equation (12).}
\end{figure}

Note that in the chiral limit, $m_0\to 0$, (12) admits 
two solutions $m=0$ or $m\ne 0$ in dependence on $GN_{{\rm c}\ >}^{\ <}
\left(GN_{\rm c}\right)_{\rm crit.}$. Thus, for large enough 
couplings we find a nonvanishing quark condensate $\left<
\bar q q\right>\sim {\rm tr}{\,} S(x,x)$ signalling spontaneous 
breakdown of chiral symmetry.

It is further convenient to perform a shift in the integration 
variables

$$ \sigma\to m+\frac{\sigma'}{\sqrt{N_{\rm c}}},$$

$$\vec \pi\to\frac{\vec\pi}{\sqrt{N_{\rm c}}}.$$
In order to obtain from the nonlocal expression (10) a local 
effective meson Lagrangian we have to apply the following 
recipes:

\begin{enumerate}
\begin{enumerate}
\item 
loop expansion of the determinant in the fields $\sigma', 
\vec\pi$, i.e.

$$N_{\rm c}{\rm tr}\ln \left\{\left(\I\hat\partial-m\right)
\left[1-\frac{1}{\I\hat\partial-m}\left(\frac{\sigma'}
{\sqrt{N_{\rm c}}}+\I\gamma_5\vec\tau\frac{\vec\pi}{\sqrt{N_{\rm c}}}
\right)\right]\right\}=$$

\begin{figure}[h]
\def\emline#1#2#3#4#5#6{%
       \put(#1,#2){\special{em:moveto}}%
       \put(#4,#5){\special{em:lineto}}}
\def\newpic#1{}

\unitlength=1mm
\special{em:linewidth 0.4pt}
\linethickness{0.4pt}
\begin{picture}(115.00,30.00)
\put(0.00,20.00){\makebox(0,0)[cc]{=}}
\put(15.00,20.00){\circle{10.20}}
\emline{20.00}{25.00}{1}{10.00}{15.00}{2}
\emline{14.00}{15.00}{3}{14.00}{10.00}{4}
\emline{16.00}{15.00}{5}{16.00}{10.00}{6}
\put(21.00,10.00){\makebox(0,0)[cc]{$\sigma'$}}
\put(25.00,20.00){\makebox(0,0)[cc]{+}}
\emline{29.00}{21.00}{7}{37.00}{21.00}{8}
\emline{29.00}{19.00}{9}{37.00}{19.00}{10}
\put(42.00,20.00){\circle{10.00}}
\emline{47.00}{21.00}{11}{56.00}{21.00}{12}
\emline{47.00}{19.00}{13}{56.00}{19.00}{14}
\put(32.00,25.00){\makebox(0,0)[cc]{$\sigma', \vec\pi$}}
\put(52.00,25.00){\makebox(0,0)[cc]{$\sigma', \vec\pi$}}
\put(60.00,20.00){\makebox(0,0)[cc]{+}}
\put(70.00,20.00){\circle{10.00}}
\emline{69.00}{25.00}{15}{69.00}{30.00}{16}
\emline{71.00}{25.00}{17}{71.00}{30.00}{18}
\emline{66.00}{17.00}{19}{61.00}{12.00}{20}
\emline{67.00}{16.00}{21}{62.00}{11.00}{22}
\emline{73.00}{16.00}{23}{78.00}{11.00}{24}
\emline{74.00}{17.00}{25}{79.00}{12.00}{26}
\put(82.00,20.00){\makebox(0,0)[cc]{+}}
\put(92.00,20.00){\circle{10.00}}
\emline{88.00}{17.00}{27}{83.00}{12.00}{28}
\emline{89.00}{16.00}{29}{84.00}{11.00}{30}
\emline{95.00}{16.00}{31}{100.00}{11.00}{32}
\emline{96.00}{17.00}{33}{101.00}{12.00}{34}
\put(88.00,23.00){\rule{0.00\unitlength}{0.00\unitlength}}
\emline{88.00}{23.00}{35}{83.00}{28.00}{36}
\emline{89.00}{24.00}{37}{84.00}{29.00}{38}
\emline{95.00}{24.00}{39}{100.00}{29.00}{40}
\emline{96.00}{23.00}{41}{101.00}{28.00}{42}
\put(105.00,20.00){\makebox(0,0)[cc]{+}}
\put(53.00,0.00){\makebox(0,0)[cc]{finite diagrams.}}
\end{picture}
\caption[]{Loop expansion of the fermion determinant.}
\end{figure}

Here we have omitted the unimportant constant term $N_{\rm c} {\rm 
tr} \ln\left(\I\hat\partial-m\right)$ and used the formula 
$\ln\left(1-x\right)=-\left[x+\frac{x^2}{2}+\cdots\right]$. 
(The tadpole diagram linear in $\sigma'$ cancels by a 
corresponding linear term arising from the first term in (10) 
due to the gap equation (12).)
\item Low-momentum expansion of loop diagrams corresponding 
to the so-called gradient expansion of meson fields in 
configuration space.
\item Field renormalization,

$$\left(\sigma',\vec\pi\right)\to Z^{\frac{1}{2}}
\left(\sigma',\vec\pi\right).$$
\end{enumerate}
\end{enumerate}
Discarding the contributions of finite $n$-point diagrams 
with $n>4$ in the loop expansion leads to a linear $\sigma$ 
model of composite fields (the prime in $\sigma'$ is now 
omitted) [3]

$$
{\cal L}_{{\rm Eff.}} \left(\sigma,\vec\pi\right)=\frac{1}{2}
\sigma\left(-\Box-m_\sigma^2\right)\sigma+\frac{1}{2}
\vec\pi\left(-\Box-m_\pi^2\right)\vec\pi-$$

\begin{equation}
-g_{\sigma\pi\pi}\sigma\left(\sigma^2+\vec\pi^2\right)-
g_{4\pi}\left(\sigma^2+\vec\pi^2\right)^2.
\end{equation}
Notice that the masses and coupling constants of mesons 
are fixed by the quark model parameters and the (finite) 
momentum cut-off $\Lambda$,

\begin{equation}
m_\pi^2=m_0\frac{g_{\pi qq}^2}{mG},~~ m_\sigma^2=m_\pi^2+
4m^2,
\end{equation}

$$g_{\pi qq}=g_{\sigma qq}=\left(\frac{Z}{N_{\rm c}}\right)^{\frac{1}{2}}=
\left\{4N_{\rm c}I_2
\right\}^{-\frac{1}{2}}=O\left(\frac{1}{\sqrt{N_{\rm c}}}\right),$$

\begin{equation}
g_{\sigma\pi\pi}=g_{3\sigma}=\frac{m}{\left(N_{\rm c}I_2
\right)^{\frac{1}{2}}},
\end{equation}

$$g_{4\pi}=g_{4\sigma}=\frac{1}{8N_{\rm c}I_2},$$
where 

$$I_2=-\I \int\limits_{}^\Lambda \frac{\D^4 k}{\left(2\pi\right)^4}\frac{1}
{\left(k^2-m^2\right)^2}.$$
Notice that in the chiral limit $m_0\to 0$, the pion becomes 
the massless Goldstone boson associated to DBCS. 
Introducing electroweak interactions at the quark level 
leads to the additional diagram shown in Fig.5.

\begin{figure}[htb]

\def\emline#1#2#3#4#5#6{%
       \put(#1,#2){\special{em:moveto}}%
       \put(#4,#5){\special{em:lineto}}}
\def\newpic#1{}

\unitlength=1mm
\special{em:linewidth 0.4pt}
\linethickness{0.4pt}
\begin{picture}(105.00,46.00)
\put(47.0,18.00){\circle{20.00}}
\emline{20.00}{19.00}{1}{40.00}{19.00}{2}
\emline{20.00}{17.00}{3}{40.00}{17.00}{4}
\put(56.50,17.50){\oval(5.00,4.50)[t]}
\put(61.50,17.50){\oval(5.00,4.50)[b]}
\put(66.50,17.50){\oval(5.00,4.50)[t]}
\put(71.50,17.50){\oval(5.00,4.50)[b]}
\put(72.00,26.00){\makebox(0,0)[cc]{$W$}}
\put(23,26){\makebox(0,0)[cc]{$\pi$}}
\end{picture}

\caption{Quark diagram describing the weak transition $\pi\to W$.}
\end{figure}
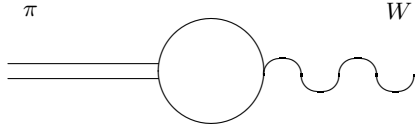

Fig.5 leads to the interaction Lagrangian 

\begin{equation}
\Delta {\cal L}=\frac{g}{2}W_\mu^i\left(-\frac{m}{g_{\pi qq}}
\partial^\mu\pi^i\right),
\end{equation}
which yields just the PCAC meson current. Here the ratio 
$m/g_{\pi qq}$ has the meaning of the pion decay constant 
$F_\pi$. We thus obtain the Goldberger-Treiman relation 

\begin{equation}
F_\pi=\frac{m}{g_{\pi qq}},
\end{equation}
valid at the quark level.

In the following subsection we shall show how one can derive the related 
nonlinear version of the $\sigma$ model.
\subsection{Nonlinear $\sigma$ model}
Nonlinear chiral meson Lagrangians have been 
introduced a long time ago 
in the literature [21-23] and are now widely used in low-energy 
hadron physics [1,24]. Let us demonstrate how they can be 
derived from a NJL quark model. To this end, it is convenient 
to use in (6) an exponential parametrization of the meson 
fields

\begin{equation}
\sigma+\I\gamma_5\vec\tau\cdot\vec\pi=\left(m+\tilde\sigma
\right)\E^{-\I\gamma_5\frac{\vec\tau\cdot\vec\varphi}
{F_\pi}},
\end{equation}
which yields
\begin{equation}
\sigma^2+\vec\pi^2=\left(m+\tilde\sigma\right)^2.
\end{equation}
Let us absorb the exponential in (19) by introducing chirally 
rotated ``constituent'' quarks $\chi$, 

\begin{equation}
q=\E^{\I\gamma_5\frac{\vec\tau\cdot\vec\xi}{2}}\chi;~~
\vec\xi=\frac{\vec\varphi}{F_\pi}
\end{equation}
leading to the Lagrangian\footnote{There arises also an 
additional anomalous Wess-Zumino term from the fermion 
measure of the path integral [4] which is discarded here.}

$$
{\cal L}_{{\rm NJL}}^{\chi{\rm M}}=-\frac{1}{2G}\left(m
+\tilde\sigma\right)^2+\Delta{\cal L}_{{\rm sb}}+$$ 
      
\begin{equation}
+\bar\chi\left[\I\gamma_\mu\left(\partial^\mu+\E^{-\I\gamma_5
\frac{\vec\tau\cdot\vec\xi}{2}}\partial^\mu\E^{\I\gamma_5
\frac{\vec\tau\cdot\vec\xi}{2}}\right)-m-\tilde\sigma\right]
\chi,
\end{equation}
with 

\begin{equation}
\Delta{\cal L}_{{\rm sb}}=\frac{m+\tilde\sigma}{16G}m_0
{\rm tr}{\,}\left(\E^{-\I\gamma_5\vec\tau\cdot\vec\xi}+
{\rm h.c.}\right)
\end{equation}
being a symmetry-breaking mass term.

Note the nonlinear transformation law of the meson field 
$\vec\xi(x)$ under global chiral transformations $g\in 
{\rm SU}(2)_{{\rm A}}\times {\rm SU}(2)_{{\rm V}}$ [22,23], 

\begin{equation}
g\E^{\I\gamma_5\frac{\vec\tau\cdot\vec\xi(x)}{2}}=
\E^{\I\gamma_5\frac{\vec\tau\cdot\vec\xi'(x)}{2}}h(x),
\end{equation}
where 

$$h(x)=\E^{\I\frac{\vec\tau}{2}\cdot\vec u\left(\vec\xi (x), 
g\right)}\in {\rm SU}(2)_{{\rm V,{\,}loc.}}$$
is now an element of a {\em local} vector group. It is 
convenient to introduce the Cartan decomposition 

\begin{equation}
\E^{-\I\gamma_5\frac{\vec\tau\cdot\vec\xi}{2}}\partial_\mu
\E^{\I\gamma_5\frac{\vec\tau\cdot\vec\xi}{2}}=\I\gamma_5
\frac{\vec\tau}{2}\cdot\vec{\cal A}_\mu\left(\vec\xi\right)+
\I\frac{\vec\tau}{2}\cdot\vec{\cal V}_\mu\left(\vec\xi\right).
\end{equation}
It is then easy to see that the fields transform under the 
local group ${\rm SU}(2)_{{\rm V,{\,}loc.}}$ as follows 
(${\cal A}_\mu\equiv\frac{\vec\tau}{2}\cdot\vec{\cal A}_\mu$, etc.) [23]

$$\chi\to\chi'=h(x)\chi$$

$${\cal V}_\mu\to {\cal V}_\mu'=h(x){\cal V}_\mu h^{\dagger} (x)-
h(x)\I\partial_\mu h^{\dagger} (x)$$

\begin{equation}
{\cal A}_\mu\to {\cal A}_\mu'=h(x){\cal A}_\mu h^{\dagger} (x).
\end{equation}
Thus, ${\cal V}_\mu$ is a gauge field with respect to ${\rm SU}(2)_{{\rm 
V, loc.}}$. This allows one to define the following 
chiral-covariant derivative of the quark field $\chi$

\begin{equation}
D_\mu \chi=\left(\partial_\mu+\I{\cal V}_\mu\right)\chi.
\end{equation}
Using (22),(25) and (27), the inverse propagator of the 
rotated $\chi$ field  takes the form

\begin{equation}
S_\chi^{-1}=\I\hat D-m-\tilde\sigma-\hat{\cal A}\gamma_5.
\end{equation}
The nonlinear $\tilde\sigma$ model is now obtained by 
``freezing'' the $\tilde\sigma$ field, performing the path 
integral over the $\chi$ field and then using again the loop 
expansion for the quark determinant $\det S_\chi^{-1}$. 
Choosing a gauge-invariant regularization, the loop diagram 
with two external ${\cal A}_\mu$ fields generates a mass-like term 
for the axial ${\cal A}_\mu\left(\vec\xi\right)$ field contributing 
to the effective Lagrangian\footnote{Note that usually 
generated field strength terms $-\frac{1}{4}{\cal A}_{\mu\nu}^2, -\frac{1}
{4}{\cal V}_{\mu\nu}^2$ vanish for Cartan fields (25). A 
mass-like term $\sim {\cal V}_\mu^2$ does not appear due to 
gauge-invariant regularization.}

\begin{equation}
{\cal L}_{{\rm nlin.}}^\sigma=\frac{m^2}{g_{\pi qq}^2}
{\rm tr}_{\rm F}{\cal A}_\mu^2+\Delta{\cal L}_{\rm sb}.
\end{equation}
Here, the symmetry-breaking term is taken over from (22), and 
the first constant term has been omitted. 

As has been shown in [23], $\vec {\cal A}_\mu$ is just the chiral-covariant 
derivative of the $\vec\xi$ field admitting the expansion 

\begin{equation}
{\cal A}_\mu^i\equiv D_\mu\xi^i=\partial_\mu\xi^i+O\left(\xi^3\right)=
\frac{1}{F_\pi}\partial_\mu\varphi^i+O\left(\varphi_i^3\right).
\end{equation}
Thus, we obtain the nonlinear $\sigma$ model

\begin{equation}
{\cal L}_{{\rm nlin.}}^\sigma=\frac{F_\pi^2}{2}
D_\mu\xi^i D^\mu\xi^i+\Delta {\cal L}_{{\rm sb}}=
\end{equation}

$$=\frac{1}{2}\vec\varphi\left(-\Box-m_\pi^2\right)\vec\varphi+
O\left(\varphi_i^3\right)$$
reproducing Weinberg's result [21].      

\section{Conclusions and Outlook}
In this talk I have shown you that the path integral bosonization approach 
applied to QCD-motivated NJL models is a powerful tool in order to 
derive low-energy effective meson Lagrangians corresponding to the 
(nonperturbative) chiral sector of QCD. 

The above considerations have further been extended to calculate 
higher-order derivative terms in meson fields by applying heat-kernel 
techniques to the evaluation of the quark determinant [4]. This allows, 
in particular, to estimate all the low-energy structure constants 
$L_i$ introduced by Gasser and Leutwyler [24]. Moreover, it is not 
difficult to include vector and axial-vector currents into the NJL 
model and to consider the chiral group $U(3)\times U(3)$. In a series 
of papers [1,3,4,25] it has been shown that the low-energy dynamics 
of light pseudoscalar, vector and axial-vector mesons is described 
surprisingly well by effective chiral Lagrangians resulting from the 
bosonization of QCD-motivated NJL models. These Lagrangians embody 
the soft-pion theorems, vector dominance, Goldberger-Treiman and KSFR 
relations and the integrated chiral anomaly.

Finally, we have investigated the path integral bosonization of an 
extended NJL model including DBCS of light quarks and heavy quark 
symmetries of heavy quarks [26] (see also [27]). This enables one 
to derive an effective 
Lagrangian of pseudoscalar, vector and axial-vector $D$ or $B$ mesons 
interacting with light $\pi, \rho$ and $a_1$ mesons. Note that the use 
of the low-momentum expansion in the evaluation of the quark determinant 
restricts here the applicability of the resulting effective Lagrangian to 
such (decay) processes where the momentum of the light mesons is 
relatively small. 

Summarizing, I thus hope to have convinced you that the path integral 
bosonization approach to QCD (in its bilocal or simpler local formulation) 
is a very interesting nonperturbative {\em analytical} method which, being 
complementary to numerical studies of lattice QCD, is worth to be 
developed further. In the next lecture, I will consider the related 
path integral ``hadronization'' of baryons.

\section*{References}

1.~For a recent review with further references see: Ebert, D., 
Reinhardt, H., and Volkov, M.K.: Progr. Part. Nucl. Phys. {\bf 33} 
(1994),1.\\
2.~Nambu, Y. and Jona-Lasinio, G.: Phys. Rev. {\bf 122} (1961), 345, 
ibid. {\bf 124} (1961), 246.\\
3.~Ebert, D. and Volkov, M.K.: Yad. Fiz. {\bf 36} (1982), 1265, Z. Phys. 
{\bf C16} (1983), 205.\\
4.~Ebert, D. and Reinhardt, H.: Nucl. Phys. {\bf B271} (1986), 188.\\
5.~Bardeen, J., Cooper, L.W., and Schriffer, J.R.: Phys. Rev. {\bf 106} 
(1957), 162.\\
6.~Bogoliubov, N.N.: Zh. Eksp. Teor. Fiz. {\bf 34} (1958), 73.\\
7.~Hubbard, J.: Phys. Rev. Lett. {\bf 3} (1959), 77.\\
8.~Stratonovich, R.L.: Sov. Phys. Dokl. {\bf 2} (1957), 416.\\ 
9.~Ebert, D. and Pervushin, V.N.: Teor. Mat. Fiz. {\bf 36} (1978), 313; 
Ebert, D. and Kaschluhn, L.: Nucl. Phys. {\bf B355} (1991), 123.\\
10.Ebert, D., Reinhardt, H., and Pervushin, V.N.: Sov. J. Part. Nucl. 
{\bf 10} (1979), 444.\\
11.Kleinert, H.: Phys. Lett. {\bf B62} (1976), 77 and Erice Lectures 
(1978).\\
12.Cahill, R.T., Praschifka, J., and Roberts, D.: Phys. Rev. {\bf D36} 
(1987), 209.\\
13.Efimov, G. and Nedelko, S.: Phys. Rev. {\bf D51} (1995), 176.\\
14.Ebert, D. and Reinhardt, H.: Teor. Mat. Fiz. {\bf 41} (1979), 139.\\
15.Azakov, S.I.: ``Two-dimensional Hubbard Model and Heisenberg 
Antiferromagnet'' (in Lecture Notes, IASBS, Zanjan (1997)).\\
16.Wolff, U.: Nucl. Phys. {\bf B225} (1983), 391.\\
17.Mandelstam, S.: Phys. Rev. {\bf D11} (1975), 3026; see also: 
Coleman, S.: Phys. Rev. {\bf D11} (1975), 2088; Luther, A. and 
Peschel, I.: Phys. Rev. {\bf B9} (1974), 2911.\\
18.Witten, E.: Comm. Math. Phys. {\bf 92} (1984), 455.\\
19.Date, G.D., Frishman, Y., and Sonnenschein, J.: Nucl. Phys. 
{\bf B283} (1987), 365; Frishman, Y. and Sonnenschein, J.: Nucl. Phys. 
{\bf B294} (1987), 801.\\
20.Giamarchi, T.: ``Bosonization in Condensed Matter Physics'', see 
these Proceedings.\\ 
21.Weinberg, S.: Phys. Rev. Lett. {\bf 18} (1967), 188.\\
22.Coleman, S., Wess, J., and Zumino, B.: Phys. Rev. {\bf 177} (1969), 
2239; Callan, G.G. et al.: Phys. Rev. {\bf 177} (1969), 2247.\\
23.Ebert, D. and Volkov, M.K.: Forschr. Phys. {\bf 29} (1981), 35.\\   
24.Gasser, J. and Leutwyler, H.: Nucl. Phys. {\bf B250} (1985), 465, 
517, 539.\\ 
25.Ebert, D., Bel'kov, A.A., Lanyov, A.V., and Schaale, A.: Int. J. 
Mod. Phys. {\bf A8} (1993), 1313.\\
26.Ebert, D., Feldmann, T., Friedrich, R., and Reinhardt, H.: Nucl. 
Phys. {\bf B434} (1995), 619; Ebert, D., Feldmann, T., and Reinhardt, 
H.: Phys. Lett. {\bf B388} (1996), 154.\\
27.Bardeen, W.A. and Hill, C.T.: Phys. Rev. {\bf D49} (1993), 409; 
Novak, M.A., Rho, M., and Zahed, I.: Phys. Rev. {\bf D48} (1993), 4370.

\end{document}